# Gauge-induced Floquet topological states in photonic waveguides

Wange Song[1,2], Hanmeng Li[1,2], Shenglun Gao[1,2], Shengjie Wu[1,2], Chen Chen[1,2], Shining Zhu[1,2], and Tao Li[1,2*]

[1]National Laboratory of Solid State Microstructures, Key Laboratory of Intelligent Optical Sensing and Integration, Jiangsu Key Laboratory of Artificial Functional Materials, College of Engineering and Applied Sciences, Nanjing University, Nanjing, 210093, China.

[2]Collaborative Innovation Center of Advanced Microstructures, Nanjing, 210093, China.

*Corresponding authors: Prof. Tao Li, Email: taoli@nju.edu.cn

**Abstract**

Tremendous efforts have been devoted to the search for exotic topological states, which usually exist at an interface between lattices with differing topological invariants according to the bulk-edge correspondence. Here, we show a new finding of topological states localized at the interface between two gauge-shifted Floquet photonic lattices, despite the same topological order across the entire structure. The quasienergy band structures reveal that these new interface modes belong to the Floquet $\pi$ modes, which are further found to enable a robust one-way propagation thanks to the flexible control of the Floquet gauge. The intriguing propagations of these $\pi$-interface modes are experimentally verified in a silicon waveguides platform at near-infrared wavelengths, which show both broad working bandwidth and high tolerance to the structural fluctuations. Our approach provides a new route for light manipulations with robust behaviours in Floquet engineering and beyond.

**Introduction**

Optical topological edge state (TES) has exhibited robustness against local structural fluctuations and disorders that significantly plays an important role in signal transport (*1-3*). It has been demonstrated in various systems, including quasicrystals (*4-6*), high-order topological insulators (*7-10*), non-Hermitian systems (*11-13*), and periodically driven Floquet systems (*14-20*). Among them, a common notion is that a topological state is expected to form at an interface between two systems with different topological invariants, which is known as the famous bulk-edge correspondence (*21*). However, is it possible to create the topological interface state despite the same topological order? There are indeed some attempts to create such a topological state that is enabled by the quantum phase transition through non-Hermitian lattice engineering (*12*). Then, the question remains in the Hermitian system.

Artificial gauge fields, as an important concept in physics, govern the effective dynamics of neutral particles (i.e., photons). It can be generated by properly engineering a physical system through the geometric design or external modulations, which allows us to endow systems with a wide range of intriguing features and novel functions, such as, effective magnetic field for photons (*22, 23*), photonic Aharonov-Bohm Effect (*24*), photonic topological insulators (*25-27*), dynamic localization and self-imaging (*28-31*), light guiding (*32, 33*), and non-reciprocal devices (*34, 35*). Recently, the notion of topological phases has been extended to Floquet systems where the Hamiltonian is periodic in time, $H(t+T)=H(t)$, with $T$ is the driving period (*14-20*). Periodic driving provides a powerful tool to engineer the quasienergy band structure and explore unconventional effects by tuning the driving frequency, amplitude, and the Floquet gauge. The topological properties of Floquet systems are much richer than static systems, such as anomalous topological

Floquet edge modes (*36-39*), "0" and "π" Majorana modes (*40-42*), and interacting topological Floquet phases (*43,44*).

Motivated by these considerations, here we demonstrate a radically different realization of topological states at the interface between two lattices with the same topological order but different Floquet gauges in optical silicon waveguides. Moreover, the gauge-induced π modes are found to sensitive to the initial gauge, which gives rise to an interesting one-way propagation by careful Floquet gauge engineering. Inheriting the topological protection property, these new π-interface modes and one-way propagation show good robustness against the structure fluctuations and broad working bandwidth, suggesting application potentials in photonic integrations.

**Results**

**Photonic π-interface modes induced by Floquet gauge transition**

The basic idea of gauge-induced topological mode can be illustrated by a simple example of a one-dimensional (1D) SSH model with Floquet engineering (*38, 45-48*), as the schematic depicted in Fig. 1A. Every other waveguide is periodically bent along their propagating direction $z$, *i.e.*, $x_0(z)=A\cos(\omega z+\varphi)$, where $z$ acts as the synthetic time dimension (*25*), $A$ and $\omega$ ($\omega \equiv 2\pi/P$, $P$ is the period) denote the amplitude and frequency of the sinusoidal bending, and $\varphi$ is the initial phase determined by the starting "time" $z=0$, *i.e.*, Floquet gauge. Two arrays with different Floquet gauge ($\varphi_1=\varphi_0$ at left and $\varphi_2=\varphi_1+\Delta\varphi$ at right, shown in Fig. 1A) are combined and form an interface as $\Delta\varphi\neq0$. We note that the two arrays have the same topological phase defined by a Z-valued invariant $G_\pi$, despite the change of Floquet gauge (see Fig. 1, E and F, and the supplementary materials S1). The waveguides on both sides of the interface are marked by I and II. Here, we need to mention that our Floquet modulation is not based on the commonly used purely-curved waveguides (*38*, *39*), but a straight and curved alternating arrangement. It provides a continuous tuning of the Floquet gauge difference ($\Delta\varphi$) across the interface with exact sine/cosine function, because the interface is composed of a pair of straight and curved waveguides (*i.e.*, I and II indicated in Fig. 1A). It guarantees a well-defined periodic coupling as required by the Floquet theory, while an interface composed of two shifted curved waveguides will break this definition and leads to more complex coupling circumstances. Our design enables us to apply flexible controls on the Floquet gauge, which is necessary to inspect possible anomalous topological effects. Through the coupled-mode theory (CMT), the waveguide array can be mapped into an effective 1D time-periodic tight-binding-approximated Hamiltonian as

$$H(z) = \sum_{n=1,3,5,\ldots}^{2N-1} \beta_n(z) a_n^\dagger a_n + \sum_{n=2,4,6\ldots}^{2N} \beta_0 a_n^\dagger a_n + \sum_{n=1}^{N-1}[c_0+(-1)^n \delta c_1(z)]a_n^\dagger a_{n+1} + \sum_{n=N}^{2N-1}[c_0+(-1)^n \delta c_2(z)]a_n^\dagger a_{n+1} + \mathrm{H.c.} \quad (1)$$

Here, $N$ is the number of waveguides of each array ($N$ is even), the total waveguides number are $2N$, $\beta_0$ is the propagation constant for the straight waveguides and $\beta_n(z)$ is the effective propagation constant for the curved waveguides which can be treated as a constant in the weak-guidance approximation (WGA) (*49*). The third and fourth terms in Eq. (1) represent couplings between nearest-neighbor (NN) waveguides for the left and right arrays with a constant (staggered) coupling strength $c_0$($\delta c(z)$). According to the WGA, the NN coupling strength mainly depends on their distance $d(z)$. In our configuration, the NN spacing $d_{1(2)}$ for left (right) arrays (center-to-center distance) $d_{1(2)}(z)=d_0\pm A\cos(\omega z+\varphi_{1(2)})$, where $d_0$ is the spacing without bending. Consequently, $c$ can be approximated as $c_{1(2)}(z)=c_0\pm\delta c_{1(2)}\cos(\omega z+\varphi_{1(2)})$. Note that the staggered term $\delta c_{1(2)}(z)$ is periodically modulated, and the Hamiltonian $H(z)$ in Eq. (1) thus exactly mimics the periodically driven SSH model with tunable time-periodic NN couplings and Floquet gauge modulations. According to the Floquet theory, the evolution of our system with Hamiltonian $H(z)$ is governed by the time evolution operator $U(z)=\hat{T}e^{-i\int_0^z H(z')dz'}$, where $\hat{T}$ denotes the time-ordering operator. The Floquet operator is defined as the time evolution operator for one full period $P$, given by $U(P)$ (*14*), from which a time-averaged effective Hamiltonian can be defined as $H_{\text{eff}}=(i/P)\ln U(P)$ (*14*). The eigenvalues of $H_{\text{eff}}$ correspond to the quasienergy spectrum of the system.

We first consider $\Delta\varphi=0$, in which the array as a whole exhibits no Floquet gauge transition. Figure 1C shows the quasienergy spectrum of 80 waveguides under open-boundary conditions. It is evident that the $\pi$ edge modes appear within the range $1/3 < \omega/4c_0 < 1$. As an example, the field distribution of the $\pi$ modes with $\omega/4c_0=0.4$ is shown in the inset figures, which have localized fields on the boundaries. However, when Floquet gauges of the two arrays are different, especially, anti-phased ($\Delta\varphi=\pi$), a new pair of $\pi$ modes would emerge with localized field at the interface. Figure 1B displays the quasienergy spectrum at $\omega/4c_0=0.4$ as a function of gauge difference $\Delta\varphi$, where two new discrete modes stand out from bulk modes and gradually turn into new $\pi$ modes as $\Delta\varphi$ reaches $\pi$, while the original $\pi$ modes keep unchanged. Figure 1D shows the quasienergy spectrum with $\Delta\varphi=\pi$, and $\pi$ modes also appear within the range $1/3 < \omega/4c_0 < 1$. After careful observation, one indeed finds an additional pair of $\pi$ modes in this range (see the zoom-in figure for the $\pi$ modes), and the new $\pi$ modes have strong localizations at the gauge transition interface, as shown in the inset of Fig. 1D. Without loss of generality, $\varphi_0$ is set as 0 here, which is not crucial to the existence of the new $\pi$ modes (supplementary materials S1). Figure 1 (E and F) shows the calculated topological invariant $G_\pi$ and Floquet gauge $\varphi$ across the waveguide array for $\Delta\varphi=0$ and $\Delta\varphi=\pi$, respectively (see supplementary materials S1). It is evident that the new pair of $\pi$-interface modes emerge with Floquet gauge transition, while the topological invariant keeps the same (Fig. 1F).

**Simulation and experimental results**

To verify the emergence of π-interface modes, we then investigate the dynamics of light with single waveguide excitation on waveguide-I at the interface. Figure 2 (*D* to *F*) shows the theoretical (CMT) results (left panels) corresponding to different gauge difference $\Delta\varphi$ with a fixed $\varphi_0=0$. At $\Delta\varphi=0$ without gauge transition, the optical fields spread out into the whole lattices, indicating no localized modes at the interface. However, as the $\Delta\varphi$ increases, the optical fields gradually tends to be preferably localized at the interface and finally get trapped at the interface at $\Delta\varphi=\pi$. It is evident that the π-interface modes arise with the Floquet gauge transition. Afterwards, we carried out full-wave simulations (COMSOL MULTIPHYSICS 5.2) and experiments in a silicon waveguide array on sapphire substrate. The structural parameters of waveguide width (*w*) and height (*h*) are optimized as *w*=400 nm, *h*=220 nm to support only one fundamental mode in the silicon waveguide at $\lambda$=1550 nm with a propagation constant $\beta_0=2.1601k_0$ ($k_0$ is the free space k-vector). To realize the periodically driven condition, we consider that the silicon waveguide is sinusoidally curved along the propagation direction *z* as $x_0(z)=A\cos(2\pi z/P+\varphi)$, where *A*=71.5 nm, *P*=48.4 μm. The spacing of neighboring waveguides without bending $d_0$=618.5 nm. Based on these designs, the coupling coefficient approximately follows $c(z)=c_0\pm\delta c\cos(2\pi z/P+\varphi)$, where $c_0$=0.0811 μm$^{-1}$, $\delta c$=0.0405 μm$^{-1}$, and $\omega$=0.130 μm$^{-1}$, corresponding to the theoretical calculations. The schematics of the silicon waveguide array and the zoom-in input ends for different $\Delta\varphi$ are presented in Fig. 2 (A and B), respectively. Figure 2 (D to F) (middle panels) shows the simulated optical field evolutions of different $\Delta\varphi$ for 20 waveguides and 100 μm propagations, with respect to the CMT calculation indicated by dashed boxes in the left panels. Since the full-wave simulation is quite time consuming, we didn't perform simulations over a same large scale as the CMT calculations. Nevertheless, the simulated results are in extremely good agreement with the CMT results and clearly demonstrate the gauge transition induced π-interface modes.

The experimental samples were fabricated by E-beam lithography and inductively coupled plasma (ICP) etching process (see Materials and Methods), which include the waveguide array (80 waveguides with 200 μm length, the same as the CMT calculations) and input grating coupler that is connected to the interface waveguide. As an example, the scanning electron microscopy (SEM) images of the fabricated structure of $\varphi_0=0$, $\Delta\varphi=\pi$ are shown in Fig. 2C. In experiments, the light was input into waveguide lattice by focusing the laser (λ=1550 nm) via a grating coupler and a tapered waveguide. The transmitted signals can be collected from the scattered light from the output end. The coupling-in and out processes are imaged by a near-infrared CCD camera (Xenics Xeva-1.7-320). Figure 2 (D to F) (right panels) shows the experimental captured optical signals as light scattered from the output of the arrays with $\Delta\varphi=0$, $\pi/2$, and π respectively, corresponding to the theoretical designs. As the gauge difference $\Delta\varphi$ increases from 0 to π, the distribution of scattered light at the output gradually get localized, and eventually to a single spot at the center, evidently indicating the emergence of localized π-interface modes at $\Delta\varphi=\pi$, which well confirms the CMT and simulation results. We provided the explanation of the formation of the gauge-induced π modes by analyzing the eigenmode property of π modes in supplementary materials S2. It should be noted here that for convenience we only input the light through a single waveguide, which does not strictly conform to the eigenmodes profile of the π interface states. However, theoretical calculations show that this single waveguide input can basically excite this local interface mode as compared with the exact π mode preparations (see supplementary materials S3 for more details), and does not affect the demonstration of the fundamental physics.

**Influence of initial gauge on the π modes excitation**

Next, we fixed the gauge difference at $\Delta\varphi=\pi$ and change the initial gauge $\varphi_0$ ($\varphi_0=\pi/2$ and $\pi$). Figure 3 (A, E) shows the zoom-in schematics of input ends for $\varphi_0=\pi/2$ and $\pi$. For $\varphi_0=\pi/2$ case, there is a bright light spot at the output end that rightly locates at the interface, ensuring the emergence of $\pi$-interface mode (see Fig. 3D). Surprisingly, for another $\varphi_0=\pi$ case, we observed an obvious spreading of light at the output end (see Fig. 3H) without the $\pi$-mode localization. Both experiments are in good coincidence with the CMT (Fig. 3, B and F) and simulation results (Fig. 3, C and G). To explain it, we analyze the eigenmode property of $\pi$-interface modes at the initial stage ($z=0$), as representative distributions shown in Fig. 3 (I and J) for $\varphi_0=\pi/2$ and $\pi$, respectively. It is found that localized $\pi$-interface modes do exist for both cases, but have different mode profiles, where at the input waveguide-I (*i.e.,* waveguide #40, marked by dashed red circles) has strong field for $\varphi_0=\pi/2$ while none for $\varphi_0=\pi$. This is also true for other $\pi$-interface modes (see supplementary materials S4). Particularly, we plotted out the intensity of one of the $\pi$ modes at waveguide-I as a function of $\varphi_0$ as shown in Fig. 3K. It is evident that $\pi$ modes have the strongest field at the input waveguide for $\varphi_0=\pi/2$ but zero at $\varphi_0=\pi$. Therefore, the input from waveguide-I can excite the $\pi$ modes to the most extent in the case of $\varphi_0=\pi/2$ and give rise to robust localization, but not as $\varphi_0=\pi$ and result in a dispersive feature. This is exactly the experiments have confirmed in Fig. 3 (D and H). Thus, the initial gauge $\varphi_0$ influences the excitation condition of the $\pi$-interface modes significantly, and more detailed discussions and experimental verifications are provided in supplementary materials S5.

**Robust one-way propagation by Floquet gauge engineering**

Inspired by the emergence of gauge-induced $\pi$-interface modes and flexible control with gauge modulation, we further explore an intriguing function of robust one-way propagation. Specifically, we design a waveguide array with one end to have $\varphi_0=\pi/2$

(marked by end-A) while the opposite is $\varphi_0=\pi$ (marked by end-B), as schematically shown in Fig. 4A. We input light from both ends to examine their propagation properties. Figure 4 (B to D) shows the CMT calculated, simulated, and experimental results of field evolutions from end-A (forward), respectively. It is evident that the light propagates along the interface with topologically protected localization according to π-interface modes excitation. However, for the backward case (input from the opposite end-B), no topological mode can be excited and the light will spread out into the entire lattices, as well verified by theory, simulation, and experiment in Fig. 4 (E to G). This function is difficult to achieve in static systems, where one only realizes either the topological transports from both ends or none (*50*). Here, we definitely show the feasibility in Floquet gauge system.

In addition, to further confirm the topological protection of this π-interface mode and examine the robustness of the one-way propagation, we fabricated controlled samples with random structural discrepancies for comparisons. The theoretical analyses and experimental results are provided in supplementary materials S6. It is obvious that for forward propagation, the optical field still propagates along the boundary, which indeed suggests its robustness against disorders. Moreover, we also find that the one-way propagation is considerably insensitive to the wavelengths that indicates a broadband property. The experimentally measured contrast ratio of forward and backward propagations reaches ~98.7% (-0.059 dB) for central wavelength of 1550 nm, and has a broad bandwidth (~100 nm) for contrast ratio > -1dB (see supplementary materials S7).

**Discussion**

In conclusion, we have revealed a brand new type of gauge-induced Floquet interface state distinct from the conventional edge states in topological photonic lattices. This newly observed π-interface state arises from the Floquet gauge transition rather than the topological transition. Thanks to the initial gauge dependence, these π modes can be carefully engineered to access a one-way propagation along the interface with broad working bandwidth and robustness against structural fluctuations. The experiments were implemented in silicon waveguides platform with convincing results fully consistent with the theoretical predictions. Our finding suggests that Floquet gauge engineering does enrich new physics in topological photonics system that give rise to novel optical phenomena and functionalities inaccessible in the static systems.

**Materials and Methods**

**Sample fabrication**

The waveguide arrays and grating nanostructures are fabricated using the method of electron-beam lithography and ICP etching process. The substrate we used is 230 nm silicon deposition on 460 um alumina substrate, the substrates are cleaned in ultrasound bath in acetone and DI water for 10 min respectively and dried under clean nitrogen flow. Then 400 nm AR-N7520 photoresist film is spin-coated onto the substrate and baked at 85°C for 1 min. After that, the sample is exposed to electron beam in E-beam writer (Elionix, ELS-F125) and developed to form the AR-N7520 nanostructures. After that, the sample transferred into HSE Series Plasma Etcher 200 and etched with C4F8 and SF6 (the flow rates of these two types of gases are 75sccm:30sccm). After the ICP etching, the remaining AR-N7520 is removed by using an O2 plasma for 5 minutes.

**Measurement**

In optical measurements, a white light laser (Fianium Super-continuum, 4W) with the wavelength range from 400 nm to 2200 nm was used, and the wavelength was switched by a group of filters (FWHM =12 nm). The light with different wavelength was focused at the input grating by an objective lens (100×), and then coupled into the waveguide mode. The output signals can be detected by the scattering field from the output end. The coupling-in and -out processes were clearly imaged by a near-infrared camera (Xeva-1.7-320) by transmission scheme through another microscope objective (50×).

**Figures and Tables**

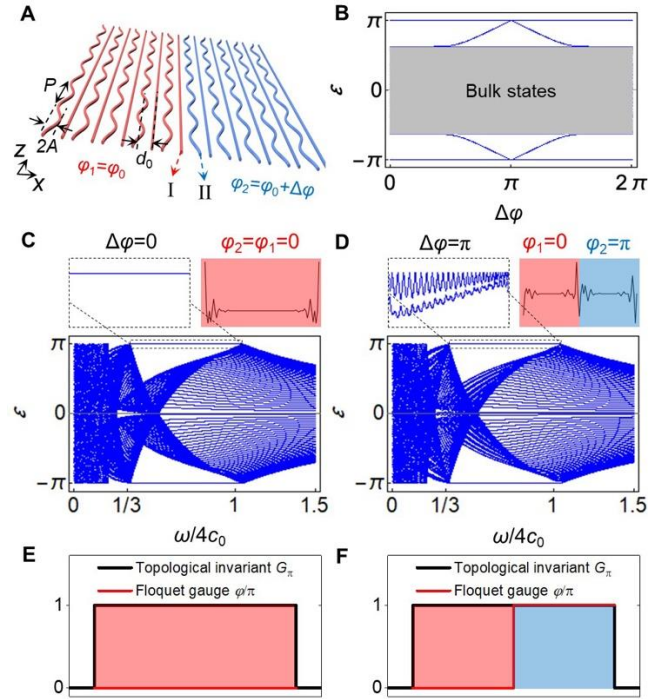

**Fig. 1. Gauge-induced photonic π-interface modes.** (**A**) Schematics of 1D periodically driven waveguide array consists of two arrays with the same topological phases but different Floquet gauge. (**B**) Corresponding quasienergies with $\omega/4c_0=0.4$ under open boundary conditions with 80 waveguides as a function of $\Delta\varphi$. At opposite Floquet gauge ($\Delta\varphi=\pi$), a new pair of π-interface modes that localized at the gauge-shifted interface emerges. (**C** and **D**) Quasienergies with 80 waveguides as a function of $\omega$ with $\Delta\varphi=0$ (C) and $\Delta\varphi=\pi$ (D). The two insets show the zoom-in of quasienergies and field distributions for one of the Floquet π modes. (**E** and **F**) Calculated topological invariant $G_\pi$ at $\omega/4c_0=0.4$ and Floquet gauge $\varphi$ across the waveguide array for $\Delta\varphi=0$ (E) and $\Delta\varphi=\pi$ (F).

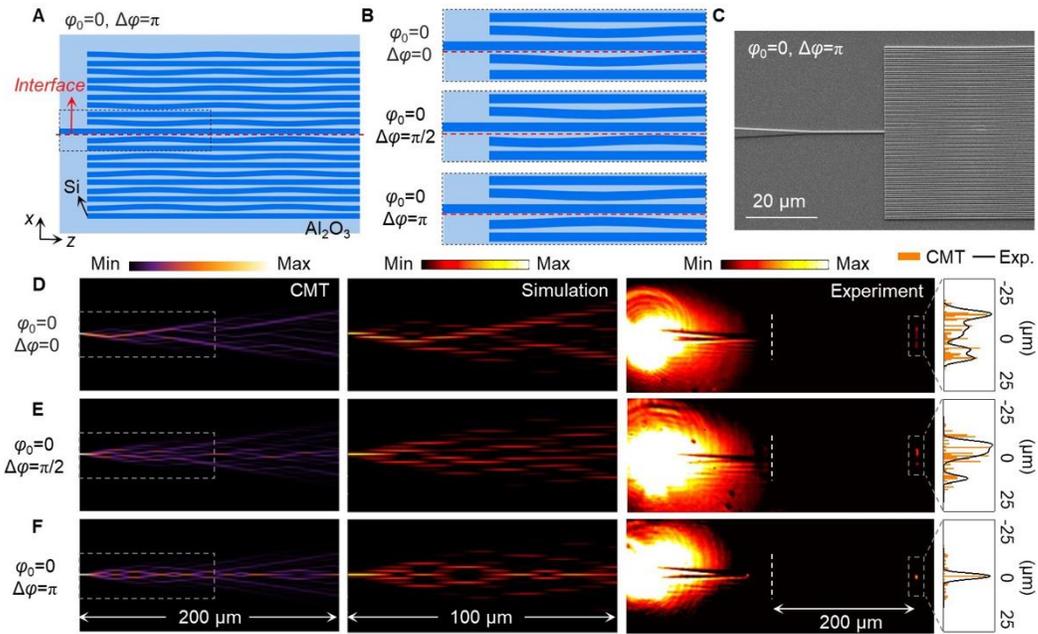

**Fig. 2. Light propagation inside waveguide arrays with the single waveguide input.**

(**A**) Schematics of the silicon waveguide array, where the interface is indicated by the red dashed line. (**B**) Zoom-in figure of the input end for different $\Delta\varphi$. (**C**) SEM image of one of the fabricated samples. (**D, E**, and **F**) Field evolutions with different $\Delta\varphi$ and fixed $\varphi_0=0$. Left panel: CMT calculated field evolutions; Middle panel: simulation results within the boxed region in left panel; Right panel: experimental results with output intensity profiles. The new π-interface modes emerge when $\Delta\varphi$ increase from 0 to π.

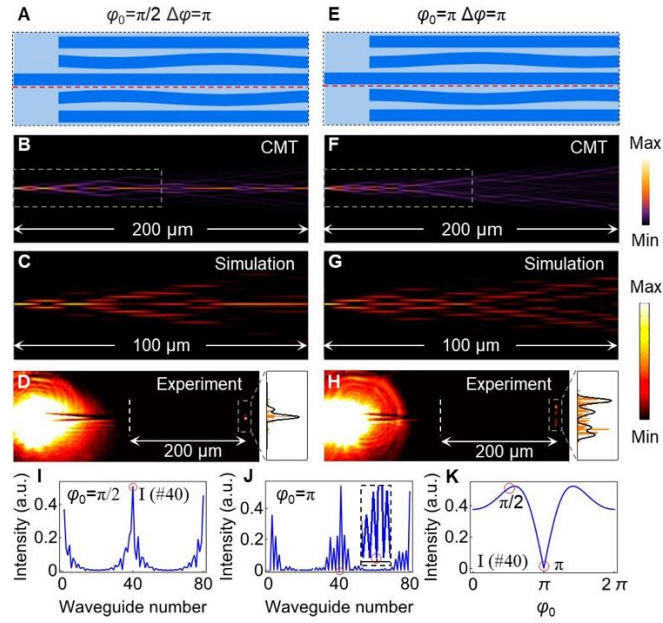

**Fig. 3 Influence of initial gauge on the π modes excitation.** (**A** to **D**), $\varphi_0=\pi/2$, $\Delta\varphi=\pi$. (**A**) Zoom-in schematic of the input end. (**B**) CMT calculated field evolution. (**C**) Simulation result within the boxed region in (B). (**D**) Experimental result with output intensity profiles. (**E** to **H**) Corresponding results for $\varphi_0=\pi$, $\Delta\varphi=\pi$. (**I** and **J**), Field distribution of the π modes at the initial stage ($z=0$) for $\varphi_0=\pi/2$ (I) and $\pi$ (J), respectively. (**K**) Field intensity of π eigenmodes at waveguide-I with respect to different initial gauge ($\varphi_0$).

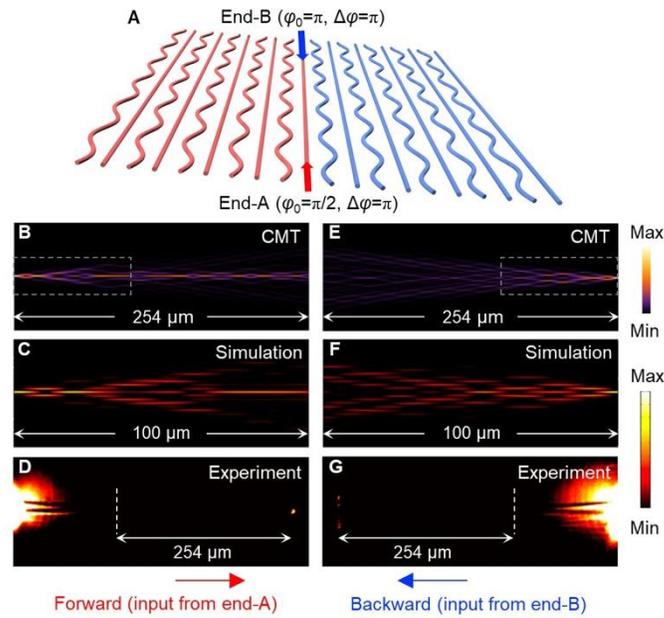

**Fig. 4 Robust one-way propagation.** (**A**) Schematics of the waveguide array with Floquet gauge modulation at the interface, where end-A and end-B represent the two ends of the waveguide array that have different initial gauge, $\varphi_0 = \pi/2$ for end-A and $\pi$ for end-B. The red and blue arrows represent input from end-A and end-B, respectively. (**B**, **C**, and **D**) CMT calculated (B), simulated (C), and experimental detected (D) field evolutions with input at end-A for the forward propagation case, showing confined interface mode. (**E**, **F**, and **G**) Corresponding results for backward propagation with input from the opposite end-B, showing dispersive feature.

**References and Notes**

1. L. Lu, J. D. Joannopoulos, M. Soljačić, Topological states in photonic systems, *Nat. Phys.* **12**, 626–629 (2016).

2. T. Ozawa, H. M. Price, A. Amo, N. Goldman, M. Hafezi, L. Lu, M. C. Rechtsman, D. Schuster, J. Simon, O. Zilberberg, I. Carusotto, Topological photonics. *Rev. Mod. Phys.* **91**, 015006 (2019).

3. W. Song, W. Sun, C. Chen, Q. Song, S. Xiao, S. Zhu, T. Li, Robust and broadband optical coupling by topological waveguide arrays. *Laser Photon. Rev.* **4**, 1900193 (2020).

4. H. Huang, F. Liu, Quantum spin Hall effect and spin bott index in a quasicrystal lattice. *Phys. Rev. Lett.* **121**, 126401 (2018).

5. M. A. Bandres, M. C. Rechtsman, M. Segev, Topological photonic quasicrystals: fractal topological spectrum and protected transport. *Phys. Rev. X* **6**, 011016 (2016).

6. Y. E. Kraus, Y. Lahini, Z. Ringel, M. Verbin, O. Zilberberg, Topological states and adiabatic pumping in quasicrystals. *Phys. Rev. Lett.* **109**, 106402 (2012).

7. W. A. Benalcazar, B. A. Bernevig, T. L. Hughes, Quantized electric multipole insulators. *Science* **357**, 61-66 (2017).

8. F. Schindler, A. M. Cook, M. G. Vergniory, Z. Wang, S. S. P. Parkin, B.A. Bernevig, T. Neupert, Higher-order topological insulators. *Sci. Adv.* **4**, eaat0346 (2018).

9. A. Hassan, F. Kunst, A. Moritz, G. Andler, E. Bergholtz, M. Bourennane, Corner states of light in photonic waveguides. *Nat. Photon.* **13**, 697-700 (2019).

10. S. Mittal, V. V. Orre, G. Zhu, M. A. Gorlach, A. Poddubny, M. Hafezi, Photonic quadrupole topological phases. *Nat. Photon.* **13**, 692-696 (2019).

11. S. Weimann, M. Kremer, Y. Plotnik, Y. Lumer, S. Nolte, K. G. Makris, M. Segev, M. C. Rechtsman, A. Szameit, Topologically protected bound states in photonic parity–time-symmetric crystals. *Nat. Mater.* **16**, 433-438 (2017).

12. M. Pan, H. Zhao, P. Miao, S. Longhi, L. Feng, Photonic zero mode in a non-Hermitian photonic lattice. *Nat. Commun.* **9**, 1308 (2018).

13. W. Song, W. Sun, C. Chen, Q. Song, S. Xiao, S. Zhu, T. Li, Breakup and recovery of topological zero mode in finite non-Hermitian optical lattices. *Phys. Rev. Lett.* **123**, 165701 (2019).

14. N. H. Lindner, G. Refael, V. Galitski, Floquet topological insulator in semiconductor quantum wells. *Nat. Phys.* **7**, 490-495 (2011).

15. Z. Gu, H. A. Fertig, D. P. Arovas, A. Auerbach, Floquet spectrum and transport through an irradiated graphene ribbon. *Phys. Rev. Lett.* **107**, 216601 (2011).

16. J. Cayssol, B. Dra, F. Simon, R. Moessner, Floquet topological insulators. *Phys. Status Solidi RRL* **7**, 101 (2013).

17. M. S. Rudner, N. H. Lindner, E. Berg, M. Levin, Anomalous edge states and the bulk-edge correspondence for periodically driven two-dimensional systems. *Phys. Rev. X* **3**, 031005 (2013).

18. Y. T. Katan, D. Podolsky, Modulated Floquet topological insulators. *Phys. Rev. Lett.* **110**, 016802 (2013).

19. D. Leykam, M. C. Rechtsman, Y. D. Chong, Anomalous topological phases and unpaired Dirac cones in photonic Floquet topological insulators. *Phys. Rev. Lett.* **117**, 013902 (2016).

20. M. S. Rudner, N. H. Lindner, Band structure engineering and non-equilibrium dynamics in Floquet topological insulators. *Nat. Rev. Phys.* **2**, 229–244 (2020).

21. X. L. Qi, Y. S. Wu, S. C. Zhang, General theorem relating the bulk topological number to edge states in two-dimensional insulators. *Phys. Rev. B* **74**, 045125 (2006).

22. K. Fang, Z. Yu, S. Fan, Realizing effective magnetic field for photons by controlling the phase of dynamic modulation. *Nat. Photon.* **6**, 782–787 (2012).

23. M. Hafezi, E. A. Demler, M. D. Lukin, J. M. Taylor, Robust optical delay lines with topological protection. *Nat. Phys.* **7**, 907–912 (2011).

24. K. Fang, Z. Yu, S. Fan, Photonic Aharonov-Bohm effect based on dynamic modulation. *Phys. Rev. Lett.* **108**, 153901 (2012).

25. M. C. Rechtsman, J. M. Zeuner, Y. Plotnik, Y. Lumer, D. Podolsky, F. Dreisow, S. Nolte, M. Segev, A. Szameit, Photonic Floquet topological insulators. *Nature* **496**, 196–200 (2013).

26. M. Hafezi, S. Mittal, J. Fan, A. Migdall, J. M. Taylor, J. M. Imaging topological edge states in silicon photonics. *Nat. Photon.* 7, 1001–1005 (2013).

27. E. Lustig, S. Weimann, Y. Plotnik, Y. Lumer, M. A. Bandres, A. Szameit, M. Segev, Photonic topological insulator in synthetic dimensions. *Nature* **567**, 356–360 (2019).

28. S. Longhi, M. Marangoni, M. Lobino, R. Ramponi, P. Laporta, E. Cianci, V. Foglietti, Observation of dynamic localization in periodically curved waveguide arrays. *Phys. Rev. Lett.* **96**, 243901 (2006).

29. A. Szameit, I. L. Garanovich, M. Heinrich, A. A. Sukhorukov, F. Dreisow, T. Pertsch, S. Nolte, A. Tünnermann, Y. S. Kivshar, Polychromatic dynamic localization in curved photonic lattices. *Nat. Phys.* 5, 271–275 (2009).

30. A. Szameit, I. L. Garanovich, M. Heinrich, A. A. Sukhorukov, F. Dreisow, T. Pertsch, S. Nolte, A. Tünnermann, S. Longhi, Y. S. Kivshar, Observation of two-dimensional dynamic localization of light. *Phys. Rev. Lett.* **104**, 223903 (2010).

31. W. Song, H. Li, S. Gao, C. Chen, S. Zhu, T. Li, Sub-wavelength self-imaging in cascaded waveguide arrays. *Adv. Photon.* **2**, 036001 (2020).

32. Q. Lin, S. Fan, Light guiding by effective gauge field for photons. *Phys. Rev. X* **4**, 031031 (2014).

33. Y. Lumer, M. A. Bandres, M. Heinrich, L. J. Maczewsky, H. Herzig-Sheinfux, A. Szameit, M. Segev, Light guiding by artificial gauge fields. *Nat. Photon.* **13**, 339–345 (2019).

34. L. D. Tzuang, K. Fang, P. Nussenzveig, S. Fan, M. Lipson, Non-reciprocal phase shift induced by an effective magnetic flux for light. *Nat. Photon.* **8**, 701–705 (2014).

35. K. Fang, S. Fan, Controlling the flow of light using the inhomogeneous effective gauge field that emerges from dynamic modulation. *Phys. Rev. Lett.* **111**, 203901 (2013).

36. S. Mukherjee, A. Spracklen, M. Valiente, E. Andersson, O. Ohberg, N. Goldman, R. R. Thom- son, Experimental observation of anomalous topological edge modes in a slowly driven photonic lattice. *Nat. Commun.* **8**, 13918 (2017).

37. L. J. Maczewsky, J. M. Zeuner, S. Nolte, A. Szameit, Observation of photonic anomalous Floquet topological insulators. *Nat. Comm.* **8**, 13756 (2017).

38. Q. Cheng, Y. Pan, H. Wang, C. Zhang, D. Yu, A. Gover, H. Zhang, T. Li, L. Zhou, S. Zhu, Observation of anomalous π modes in photonic Floquet engineering. *Phys. Rev. Lett.* **122**, 173901 (2019).

39. J. Petráček, V. Kuzmiak, Dynamics and transport properties of Floquet topological edge modes in coupled photonic waveguides. *Phys. Rev. A* **101**, 033805 (2020).

40. L. Jiang, T. Kitagawa, J. Alicea, A. R. Akhmerov, D. Pekker, G. Refael, J. Ignacio Cirac, E. Demler, M. D. Lukin, P. Zoller, Majorana fermions in equilibrium and in driven cold-atom quantum wires. *Phys. Rev. Lett.* **106**, 220402 (2011).

41. A. Kundu, B. Seradjeh, Transport signatures of Floquet Majorana fermions in driven topological superconductors. *Phys. Rev. Lett.* **111**, 136402 (2013).

42. D. T. Liu, J. Shabani, J. A. Mitra, Floquet Majorana zero and $\pi$ modes in planar Josephson junctions. *Phys. Rev. B* **99**, 094303 (2019).

43. A. C. Potter, T. Morimoto, A. Vishwanath, Classification of interacting topological Floquet phases in one dimension. *Phys. Rev. X* **6**, 041001 (2016).

44. D. V. Else, C. Nayak, Classification of topological phases in periodically driven interacting systems. *Phys. Rev. B* **93**, 201103 (2016).

45. W. P. Su, J. R. Schrieffer, A. J. Heeger, Solitons in polyacetylene. *Phys. Rev. Lett.* **42**, 1698 (1979).

46. J. K. Asbóth, B. Tarasinski, P. Delplace, Chiral symmetry and bulk-boundary correspondence in periodically driven one-dimensional systems. *Phys. Rev. B* **90**, 125143 (2014).

47. V. Dal Lago, M. Atala, L. E. F. Foa Torres, Floquet topological transitions in a driven one-dimensional topological insulator. *Phys. Rev. A* **92**, 023624 (2015).

48. M. Fruchart, Complex classes of periodically driven topological lattice systems *Phys. Rev. B* **93**, 115429 (2016).

49. I. L. Garanovich, S. Longhi, A. A. Sukhorukov, Y. S. Kivshar, Light propagation and localization in modulated photonic lattices and waveguides. *Phys. Rep.* **518**, 1–79 (2012).

50. A. Blanco-Redondo, I. Andonegui, M. J. Collins, G. Harari, Y. Lumer, M. C. Rechtsman, B. J. Eggleton, M. Segev, Topological optical waveguiding in silicon and the transition between topological and trivial defect states. *Phys. Rev. Lett.* **116**, 163901 (2016).

51. M. Rodriguez-Vega, B. Seradjeh, Universal fluctuations of Floquet topological invariants at low frequencies. *Phys. Rev. Lett.* **121**, 036402 (2018).

52. A. Patsyk, U. Sivan, M. Segev, M. A. Bandres, Observation of branched flow of light. *Nature* **583**, 60–65 (2020).

53. P. W. Anderson, Absence of diffusion in certain random lattices. *Phys. Rev.* **109**, 1492–1505 (1958).

54. A. Schreiber, K. N. Cassemiro, V. Potoček, A. Gábris, I. Jex, Ch. Silberhorn, Decoherence and disorder in quantum walks: from ballistic spread to localization. *Phys. Rev. Lett.* **106**, 180403 (2011).

**Acknowledgments**

**Funding:** The authors acknowledge the financial support from the National Key R&D Program of China (2017YFA0303701, 2016YFA0202103), National Natural Science Foundation of China (Nos. 91850204, 11674167, 11621091). Tao Li thanks the support from Dengfeng Project B of Nanjing University.

**Author contributions:** T.L. and W.S. developed the idea. W.S. proposed the theoretical design, performed the numerical simulation; H.L. and S.G. fabricated the samples; W.S. performed the optical measurement with the help of S.W. and C.C.; T.L. supervised the project. W.S. and T.L. analysed the results with help from all authors. All authors discussed the results. W.S. and T.L. wrote the manuscript with input from all authors.

**Competing interests:** The authors declare that they have no competing interests.

**Data and materials availability:** All data needed to evaluate the conclusions in the paper are present in the paper and/or the Supplementary Materials. Additional data related to this paper may be requested from the authors.

**Supplementary Materials**

**1. Topological invariant and quasienergy spectrums.**

In this section, we provide the calculation of the topological invariant and detailed data of the quasienergy spectrums with respect to gauge difference $\Delta\varphi$ under different driving frequency $\omega$ and initial gauge $\varphi_0$. According to the Floquet theory, a time-averaged effective Hamiltonian can be defined as $H_{\text{eff}}=(i/P)\ln U(P)$, where $P$ is the period of the sinusoidal bending, and $U(P)$ is the time evolution operator ($U(z)=\hat{T}e^{-i\int_0^z H(z')dz'}$, where $\hat{T}$ denotes the time-ordering operator) for one full period. Both $U(z)$ and $H_{\text{eff}}$ can be Bloch decomposed as $U(z)=\prod_k U(z,k)$ and $H_{\text{eff}}=\sum_k H_{\text{eff}}(k)$, respectively, due to the translation symmetry (*14*, *38*, *48*). A Z-valued invariant can be defined for the quasienergy gap at $\pi$ for a one-dimensional (1D) periodically driven system with chiral symmetry (*38*, *48*). The $\pi$ gap invariant $G_\pi$ can be calculated through $G_\pi=\frac{i}{2\pi}\int_{-\pi}^{\pi} tr((V_\pi^+)^{-1}\partial_k V_\pi^+)dk$, where $V(z,k)\equiv U(z,k)e^{iH_{\text{eff}}(k)z}$, and $V_\pi^+$ is obtained from $V(z,k)$ at half period $V(P/2,k)=\begin{pmatrix} V_\pi^+ & 0 \\ 0 & V_\pi^- \end{pmatrix}$. The driving frequency $\omega$ determines the topological phases: $\omega/4c_0 > 1$ gives rise to $G_\pi=0$ and corresponds to the topologically trivial phase, whereas $1/3 < \omega/4c_0 < 1$ gives rise to $G_\pi=1$ and corresponds to the nontrivial phase that supports topological $\pi$ modes. For $\omega/4c_0<1/3$, the complex low-frequency behavior may result in non-integer values of the topological invariant (*51*). In the main text, the left and right arrays have the same driving frequency $\omega/4c_0=0.4$, so they have the same topological phases, and no topological mode would be expected at the interface. However, the Floquet gauge of periodically driven systems provides us a new control mechanism of the topological modes: the gauge difference $\Delta\varphi$ can affect the quasienergy band and is responsible for the emergence of $\pi$-interface modes. Figures S1(a-c) show the quasi-energy spectrums as a function of $\Delta\varphi$ with different driving frequency, i.e. with $\omega/4c_0=0.2$, 0.8, and 1.2, respectively. For the driving frequency out of the range (e.g. $\omega/4c_0=0.2$ or $\omega/4c_0=1.2$), no topological mode exists and the gauge difference $\Delta\varphi$ only slightly affect the bulk band (see Figs. S1(a) and S1(c)). However, if the driving frequency in the range (e.g. $\omega/4c_0=0.8$), the quasienergy spectrum can be drastically altered by the gauge difference $\Delta\varphi$, which can turn the bulk mode into the topological one: the π-interface modes appear at $\Delta\varphi=\pi$ (see Fig. S1(b)). Hence, the gauge can create new topological π modes that have localizations at the gauge transition interfaces within frequency range $1/3 < \omega/4c_0 < 1$.

Moreover, we change the initial gauge $\varphi_0$ to study the influence on the quasi-energy spectrums, as shown in Figs. S1(d-f), with $\varphi_0=0, \pi/2$, and π, respectively. For different $\varphi_0$, the π-interface modes always appear at $\Delta\varphi=\pi$ regardless of $\varphi_0$. It is evident that the emergence of π-interface modes has nothing to do with the initial gauge $\varphi_0$. However, $\varphi_0$ can affect the π-interface mode distributions, as will be shown in S4.

**2. Floquet modes distributions with different gauge modulations.**

In this section, we provide detailed data of the field distributions of the first four Floquet modes with the largest quasienergy $|\varepsilon|$ marked by circle, box, triangles and inverted triangles in Fig. S1, which include original π modes, bulk modes, and π-interface modes under different $\varphi_0$ and $\Delta\varphi$ parameters. Figures S2(a)-S2(d) correspond to $\Delta\varphi=0, \pi/2, 3\pi/4$, and π with $\varphi_0=0$. At $\Delta\varphi=0$, there are two π modes ($\varepsilon=\pm3.14159$, marked by circle and inverted triangles) with localized fields at two boundaries, and the other two are bulk modes ($\varepsilon=\pm1.95822$, marked by box and triangles) with extended fields (see Fig. S2(a) and Fig. S1(d) red marks). As $\Delta\varphi$ increases to $\pi/2$, the field of the two π modes still localize at the boundaries ($\varepsilon=\pm3.14159$, marked by circle and inverted triangles), but the other two bulk modes tend to be localized at the interface compared with $\Delta\varphi=0$ case, and their quasienergy $|\varepsilon|$ increases ($\varepsilon=\pm2.00659$, marked by box and triangles) (see Fig. S2(b) and Fig. S1(d) orange marks). As $\Delta\varphi$ increases, the quasienergys of these two bulk modes move towards high $|\varepsilon|$ thus beyond the bulk band and are located within the $\pi$ gap. For example, when $\Delta\varphi=3\pi/4$, two $\pi$ modes still localize at the boundaries ($\varepsilon=\pm3.14159$, marked by circle and inverted triangles), but the other two modes are localized at the interface ($\varepsilon=\pm2.48893$, marked by box and triangles). Though the localization, they are still not the topological modes (see Fig. S2(c) and Fig. S1(d) pink marks). Until the gauge difference $\Delta\varphi$ reaches $\pi$, their quasienergy $|\varepsilon|$ increases to $\pi$ ($\pm3.14154$, marked by box and triangles), which indicates that two new $\pi$ modes are formed with localized fields at the interface as termed as the $\pi$-interface modes (see Fig. S2(d) and Fig. S1(d) green marks). Besides, the original $\pi$ modes still exist ($\pm3.14158$, marked by circle and inverted triangles). Note that the quasienergys $|\varepsilon|$ of these two pairs of $\pi$ modes are not exactly equal to $\pi$, they slightly split because of the finite systems size ($N=80$) and coupling effect of the interface modes and edge modes (*13*). Accordingly, all the four $\pi$ modes have localized fields at both interface and boundaries (see Fig. S2(d)).

## 3. Comparison between the single waveguide input and exact π-mode preparations.

As we have demonstrated in Fig. S2(d), the gauge-induced Floquet $\pi$-modes consists of complex field distributions that are difficult to reproduce in experiments. The input from a single waveguide will inevitably excite other components from bulk modes. To check the influence, we carried out detailed CMT calculations on the light propagations with single waveguide input and exact mode profiles according to the four degenerated $\pi$-modes (marked by green circle, box, triangles, and inverted triangles in Fig. S2(d)), as the results shown in Fig. S3. It is found that the single waveguide input has considerable field confinement along the interface except for minor stray fields. It is due to the fact that the topological interface states have strong field localization at the interface that results in significant mode overlap with the input waveguide. As for the exact $\pi$-modes preparations, the localized field exhibits more evident periodic oscillation in its propagation (see Figs.

S5(b) and S5(c)), which is the typical property of the π-modes (*38*). In addition, there are also two edge π-modes localized at top and bottom edges of the array that are supported by topological transition, which can be observed by the boundary waveguide excitations[2]. In this work, we mainly focus on the gauge-induced π interface modes.

**4. Influence of the initial gauge on the π-interface modes distributions.**

As we have mentioned, though the initial gauge $\varphi_0$ has nothing to do with the emergence of π-interface mode, it can affect the π-interface mode distributions. Figure S4 shows the detailed data of the field distributions of the four π modes marked by circle, box, triangles, and inverted triangles in Figs. S1(d), S1(e), and S1(f) with fixed $\Delta\varphi=\pi$ and different $\varphi_0$ (0, $\pi/2$, and $\pi$). For $\varphi_0=0$, the fields of π modes are preferred to localized at interface waveguide I (i.e. #40 waveguide) rather than II (i.e. #41 waveguide). However, for $\varphi_0=\pi$, the field is mostly localized at waveguide II, while no field is at waveguide I. As for $0<\varphi_0<\pi$, both waveguide I and II share the π mode distributions. It is evident that as the initial gauge $\varphi_0$ increases from 0 to π, the preferred interface waveguide changes from I to II. These results clearly show that the π-interface modes exist in all cases of $\varphi_0$ as long as $\Delta\varphi=\pi$, and $\varphi_0$ can affect the distribution of π-mode.

To clearly show the influence of $\varphi_0$ on the π modes. We plot the field intensity of the four π-interface modes at interface waveguides I (i.e. #40 waveguide, see blue curve) and II (i.e. #41 waveguide, see red curve) as a function of $\varphi_0$ in Fig. S5. The π-interface modes preferred to be localized at waveguide I within $0<\varphi_0<2.45$ for the first two π modes (±3.14158) and $0<\varphi_0<0.49$ for another two π modes (±3.14154), while at waveguide II within $2.45<\varphi_0<\pi$ and $0.49<\varphi_0<\pi$. We note that at $\varphi_0=0$, no field is at interface waveguide II, and at $\varphi_0=\pi$, no field is at interface waveguide I. Thus, if the input is at interface waveguide I with $\varphi_0=\pi$, no topological π modes will be excited and the field would spread out across the whole arrays. However, if II is the input waveguide for the same structure, the π modes can be excited with localized fields at the interface, as verified by our later experiments shown in the S5.

**5. Experimental results of dynamics for different waveguide excitation.**

We have demonstrated that $\varphi_0$ can affect the distribution of π-interface modes. In this section, we investigate the dynamics of optical fields with different single waveguide excitation. Figure S6(a) shows the proportion of π-modes as a function of initial gauge $\varphi_0$ with fixed $\Delta\varphi=\pi$ for waveguide I input. It is found that the proportion of π-modes changes as $\varphi_0$ is modified. Specifically, the π-modes can be excited to the utmost extent if $\varphi_0=\pi/2$, but no π-mode will appear if $\varphi_0=\pi$. Hence, no localized propagation is found for $\varphi_0=\pi$ case (Figs. S6(b-d)). As a comparison. Figure S6(e) shows the proportion of π-modes as a function of initial gauge $\varphi_0$ for waveguide II input. In this case, the π-modes can be excited to the utmost extent at $\varphi_0=\pi$. Thus, the localization and the bright spot reemerge that verify the existence of the π-interface modes (Figs. S6(f-h)).

**6. Experimental verifications of robust one-way propagation against disorders.**

Next, we examine the robustness of the one-way propagation in the presence of disorder by introducing random coupling coefficients between neighboring waveguides $c_{0n}'=c_0(1+W/2\zeta_n)$, where $\zeta_n$ is the distributed random number in the interval $[-1, 1]$ modulating the coupling between the $n$th and $(n+1)$th waveguides, and they are independent of each other. $W$ is the fluctuation strength. We present the CMT calculated patterns and experimental results of forward and backward propagations with $W=0$, 0.1, 0.5, and 1 in Fig. S7. It is obvious that for the forward propagation (input from end-A), the optical field still propagates along the interface in the presence of disorders, which indeed suggests its robustness. As for backward propagation (input from end-B), the optical fields spreads out into the whole array but the spread gradually slows down as disorder increases. Interestingly, we note that there are three different stages featured by different dynamic patterns as the disorders increase. In very weak disorders ($W=0.1$), the dynamics are almost the same as the ideal one (Fig. S7(f)). When the disorders increase, but still at a weak level ($W=0.5$), the optical fields divide to form branches with enhanced intensity (Fig. S7(g)). This behavior may be related to the branched flow of light (*52*). If, on the other hand, strong disorder ($W=1$) is applied, one can observe that the wave packet is arrested at the interface (Fig. S7(h)) due to the Anderson localization (*53*, *54*).

**7. Experimental verifications of broadband one-way propagation.**

Moreover, we significantly find that the one-way propagation is also considerably insensitive to the wavelengths that indicates a broadband property. To verify this, further experiments were performed on the one-way propagation functionality with respect to different wavelengths (1490~1610nm) and Figures S8(a) and S8(b) show the CCD captured images of the forward/backward propagation (end-A/B inputs) for six different wavelengths, respectively. Expectedly, for the forward propagation, there are spots at the output end of the interface for all wavelengths. Note that in our experiments as the wavelength deviates from the center wavelength, the detected output intensity decreases. It is due to the input efficiency decreases for the mismatch between light and grating coupler, since all grating couplers are particularly designed for 1550 nm wavelength. Nevertheless, it does not affect the localization of the optical field. As for the backward propagation, no spot is observed and the output light distributes across the whole lattices for all the wavelengths. Figure S8(c) shows the extracted normalized output intensity for forward and backward propagations with respect to different wavelengths, which clearly demonstrates the one-way propagation functions. In addition, the experimentally measured contrast ratio of forward and backward propagations reaches ~ -0.059 dB for central wavelength of 1550 nm, and has a broad band (~100 nm) for contrast ratio > -1dB (Fig. S8(d), dashed line).

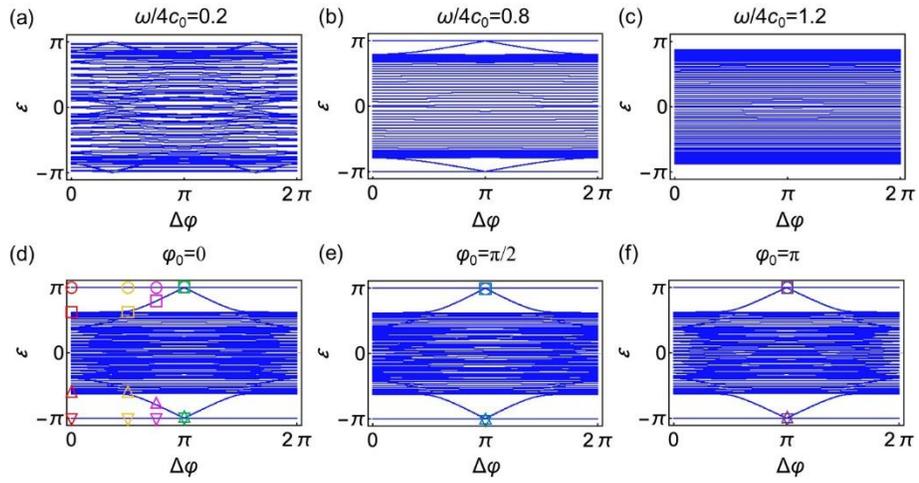

Figure S1 Quasienergy spectrums under different driving frequency $\omega$ and initial gauge $\varphi_0$. (a-c) $\omega/4c_0$=0.2, 0.8 and 1.2 with $\varphi_0$=0. (d-f) $\varphi_0$=0, $\pi/2$, and $\pi$ with $\omega/4c_0$=0.4. The circle, box, triangles and inverted triangles with different colors mark the Floquet modes shown in Figs. S2, S3 and S4.

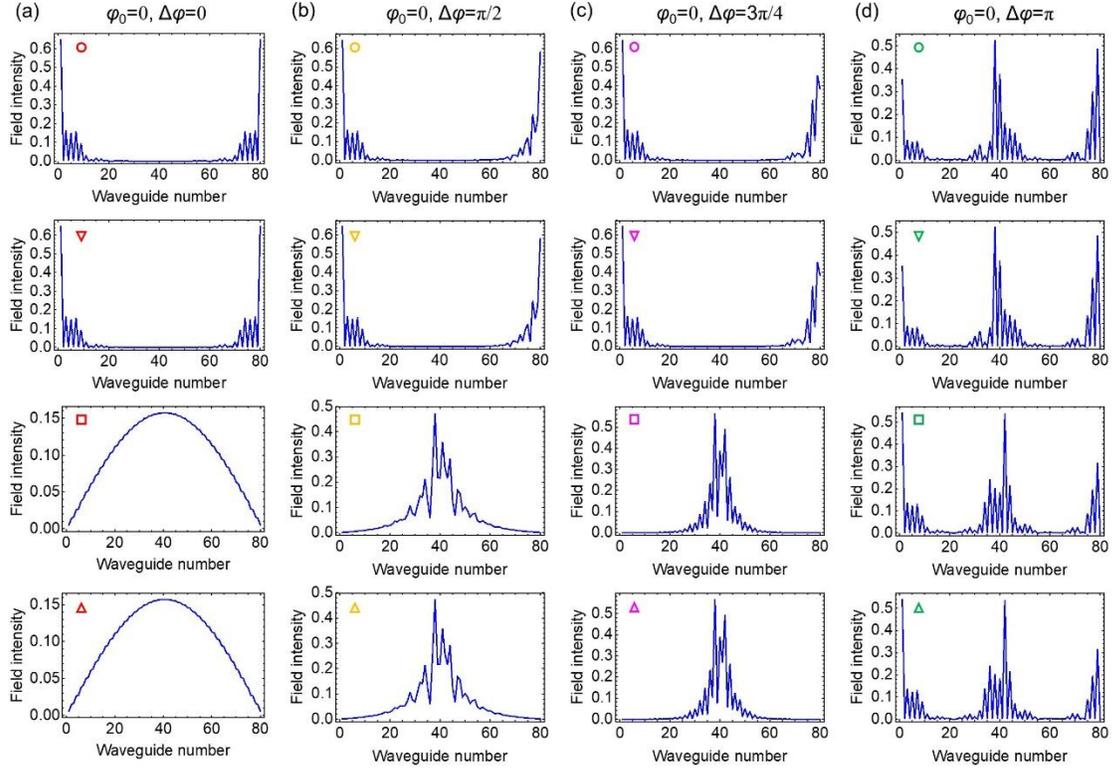

Figure S2 Field distribution of the first four Floquet modes at the initial stage ($z$=0) with the largest quasienergy $|\varepsilon|$ marked by circle, box, triangles and inverted triangles with different colors in Fig. S1(d). (a) $\Delta\varphi$=0, only two π-modes exist with localized fields on both boundaries (red circle and inverted triangles), while the other two bulk modes are extended modes (red box and triangles). (b) $\Delta\varphi$=π/2, also only two π-modes exist with localized fields on both boundaries (orange circle and inverted triangles), while the other two modes are bulk modes (orange box and triangles). Note that the two bulk modes start to get trapped at the interface. (c) $\Delta\varphi$=3π/4, also only two π-modes exist with localized fields on both boundaries (pink circle and inverted triangles), while the other two modes are localized (pink box and triangles). Note that the two localized modes are not topological modes. (d) $\Delta\varphi$=π, the four modes are all topological π-modes with localized fields at boundaries and the interface (green circle, box, triangles, and inverted triangles).

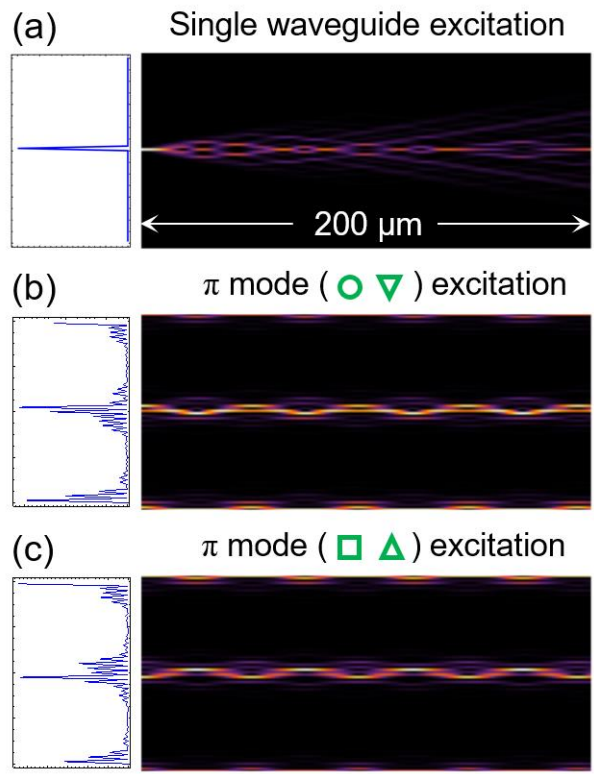

Figure S3 CMT calculated field evolutions with (a) single waveguide excitation and (b,c) Floquet π-modes excitation, the π-modes marked by green circle, box, triangles, and inverted triangles corresponds to the modes shown in Fig. S2(d).

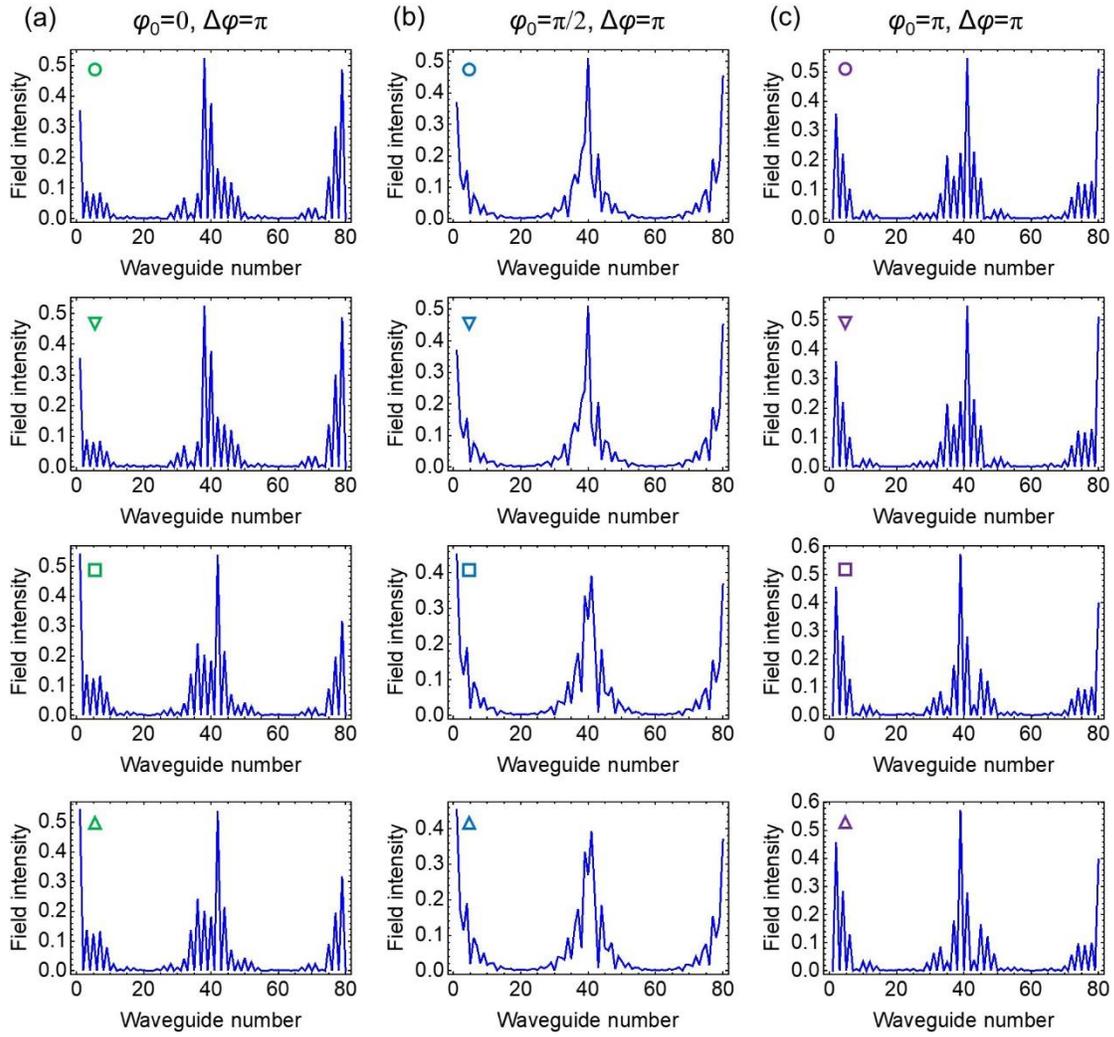

Figure S4 Field distribution of the first four Floquet modes at the initial stage ($z=0$) with the largest quasienergy $|\varepsilon|$ marked by circle, box, triangles and inverted triangles with different colors in Figs. S1(d), S1(e) and S1(f). The four modes are all topological $\pi$-modes with localized fields at boundaries and the interface. (a) $\varphi_0=0$, all the four modes have no fields at #41 waveguide, i.e. interface waveguide II. (b) $\varphi_0=\pi/2$, all the four modes have fields at both #40 and #41 waveguides, i.e. interface waveguide I and II. (c) $\varphi_0=\pi$, all the four modes have no fields at #40 waveguide, i.e. interface waveguide I.

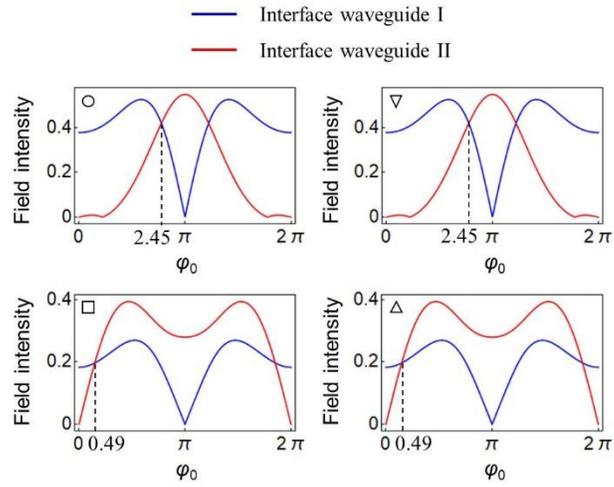

Figure S5 Field intensity of the first four Floquet π-modes at the initial stage ($z=0$) at interface waveguides I (blue curve) and II (red curve). At $\varphi_0=0$, no field is at interface waveguide II, and at $\varphi_0=\pi$, no field is at interface waveguide I.

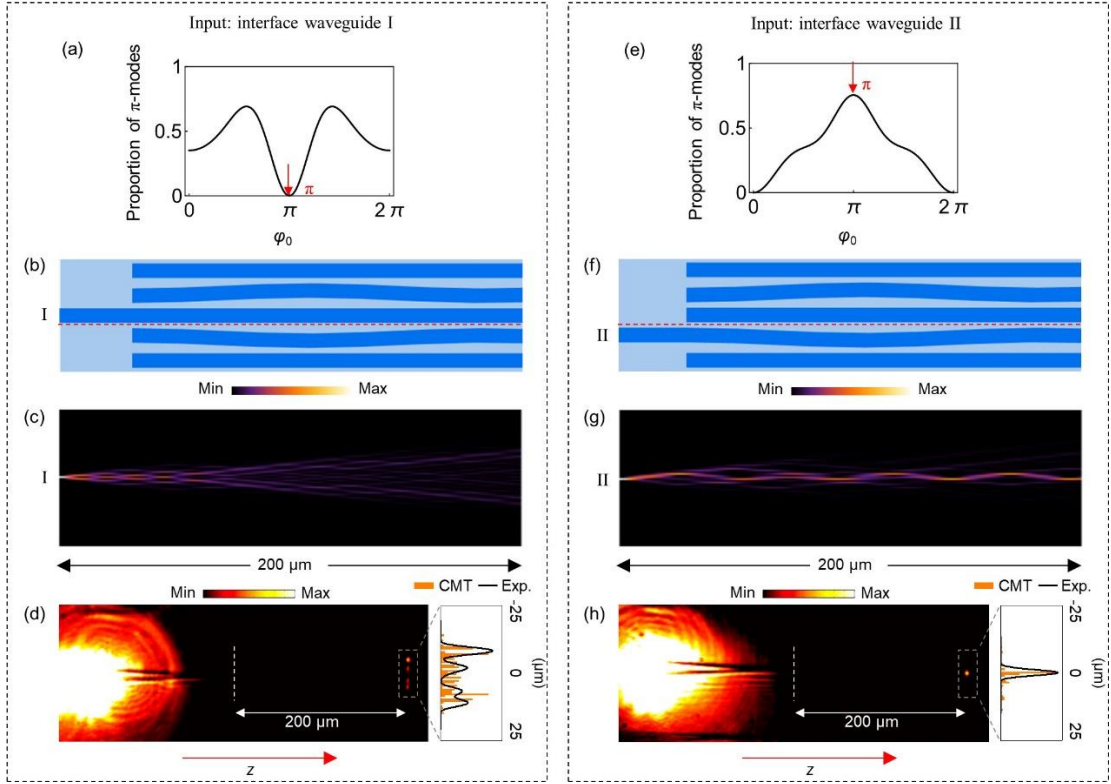

Figure S6 Experimental results of dynamics for different waveguide excitation. (a-d) Interface waveguide I input. (a) Proportion of π-modes as a function of initial Floquet gauge $\varphi_0$, the excitation of π-modes closely rely on the $\varphi_0$. (b) Enlarged pictures of the input end for $\varphi_0=\pi$, $\Delta\varphi=\pi$. (c) CMT calculated field evolutions. (d) Experimentally recorded optical propagation through the waveguide arrays and output intensity profiles. (e-h) Corresponding results for interface waveguide II input.

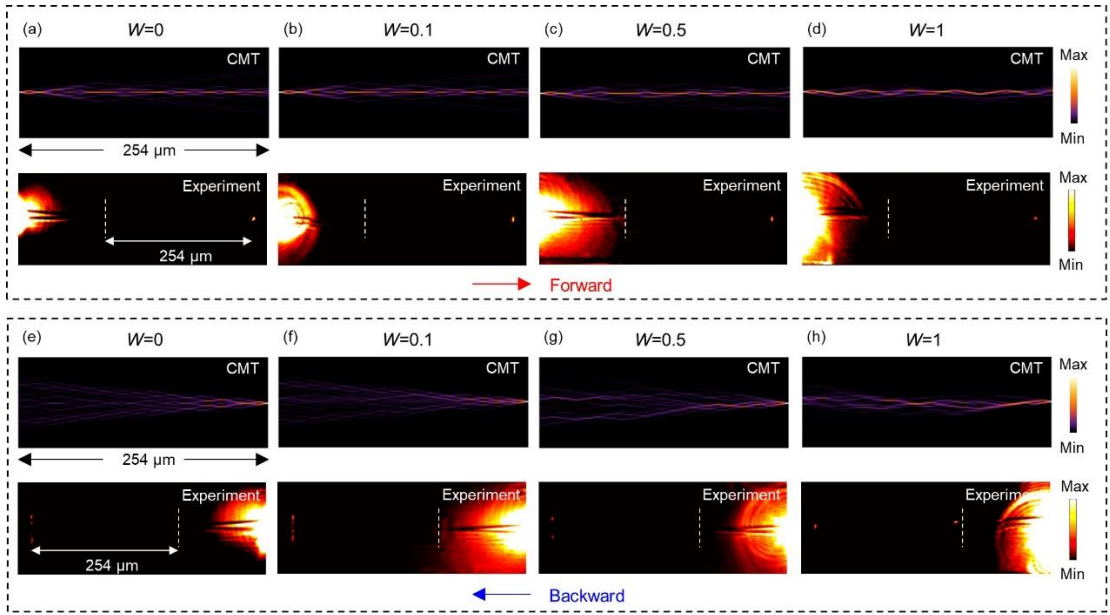

Figure S7 The dynamic evolutions of optical fields in disordered samples with different fluctuation strength $W$=0, 0.1, 0.5, and 1. (a-d) Calculated (top panels) and experimental results (bottom panels) of forward propagation. (e-h) Corresponding results for backward propagation.

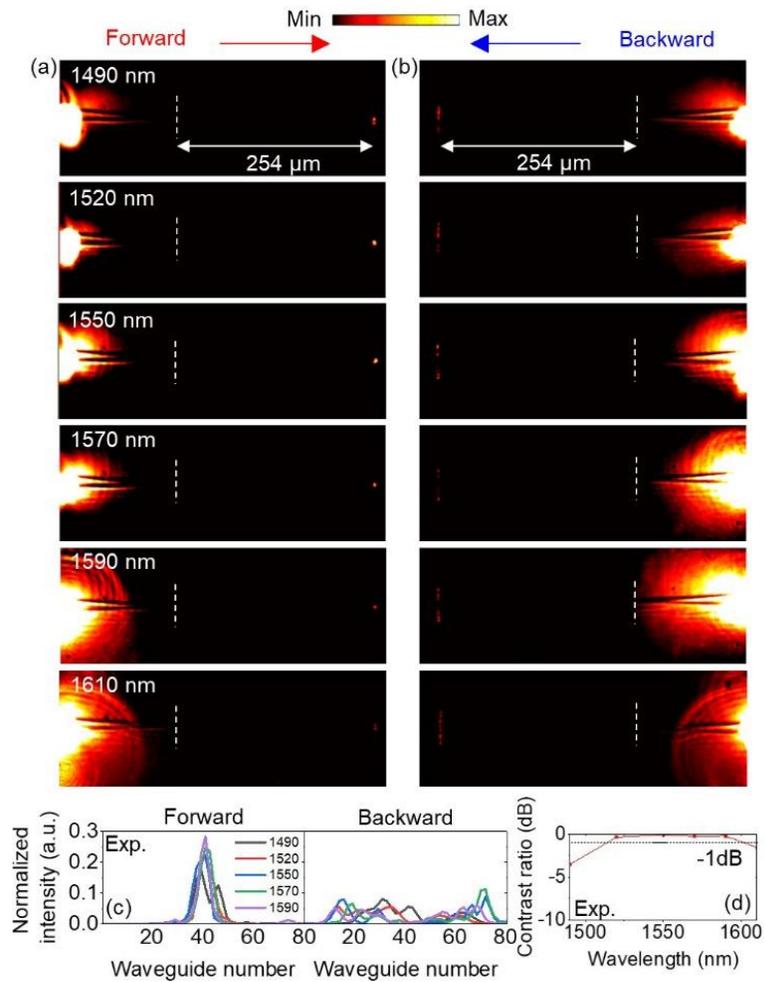

Figure S8 Broadband one-way propagations. (a,b) CCD captured images for forward (a) and backward (b) propagations. (c) Normalized output intensity profiles for different wavelengths. (d) Contrast ratio for the output field of forward and backward propagations. The intensities of three central waveguides are taken into account.